\documentstyle[twocolumn,eqsecnum,epsf,prb,aps]{revtex}
\begin{document}
\preprint{IUCM96-019}
\draft
\twocolumn[\hsize\textwidth\columnwidth\hsize\csname
@twocolumnfalse\endcsname
\title{Integer Quantum Hall Effect in Double-Layer Systems}
\author{Erik~S.~S\o rensen and A.~H.~MacDonald}
\address{Department~of~Physics, Indiana~University, Bloomington, IN~47405}  
\date{\today}
\maketitle
\begin{abstract}
We consider the localization of independent electron orbitals  
in double-layer two-dimensional electron systems
in the strong magnetic field limit.   
Our study is based on numerical Thouless number calculations for 
realistic microscopic models and on transfer matrix calculations
for phenomenological network models.  The microscopic
calculations indicate a crossover regime for weak
interlayer tunneling in which the correlation length 
exponent appears to increase.  Comparison of network 
model calculations with microscopic calculations 
casts doubt on their generic applicability.  
\end{abstract}
\vskip2pc]
%============================================================================
% BODY OF PAPER

\section{Introduction}\label{sec:intro}

The integer quantum Hall effect is generally well understood in
single-layer two-dimensional electron systems (2DES's) which are
sufficiently disordered that interactions do not play an essential role
and are in a field sufficiently strong that Landau level mixing does not
play an essential role.  In this limit, single-electron orbitals are
localized except at a critical energy $E_c$ near the center of each
disorder broadened Landau level.  For Fermi energy $E_F= E_c$,
theory~\cite{Khmel,Pruisken,Fisher,LKZ} predicts that
$(\sigma_{xx}^c,\sigma_{xy}^c) = (1,2n+1)e^2/2h$ whereas on the Hall
plateaus ($E_F \ne E_c$) $(\sigma_{xx},\sigma_{xy})=(0,n)e^2/h$ where
$n$ is the number of extended state energies below the Fermi level.   As
the critical energy is approached, the localization length for electrons
at the Fermi level is expected to have a power-law divergence, $\xi\sim
|E_F-E_c|^{-\nu}$, and $\nu$, the correlation length exponent, is
expected to be independent of $n$.  It is believed that the transition
is well described by quantum percolation~\cite{Trugman,ChalkCod} models
and semiclassical calculations~\cite{MilSok} have estimated the
correlation length exponent to $\nu=7/3$.  This picture has been
corroborated by a large number of thorough numerical
studies~\cite{ChalkDaniell,Hikami,Singh,HuoHetzelBhatt,LiuSarma} which are in
agreement with theoretical predictions for $\sigma_{xx}^c$ and
$\sigma_{xy}^c$.  Localization properties and the divergence of the
localization length have been studied extensively~
\cite{AokiAndo,Ando,AokiAndo2,Ando2,Huckestein,HuoBhatt,HuckRev,Huck3}
with the most recent estimate of the correlation length exponent being
$\nu=2.35\pm0.03$~\cite{Huck3}.  On the experimental side, measurements
of the width, $\Delta B$, of the peak of $\rho_{xx}$ as well as
$(d\rho_{xy}/dB)^{-1}$, both predicted to scale with temperature as
$T^{1/z\nu}$, yield values $1/\nu=0.42\pm0.04$\cite{HP,Koch} with $z$
assumed to be 1. Higher derivatives of $\rho_{xy}$ yields
exponents~\cite{WeiDeriv} of $n\nu$ in agreement with scaling theories
of the transition between Hall plateaus.   (Experiments in the
fractional quantum Hall regime find similar values~\cite{EngelWeiFrac}
for this exponent.) Recently the dynamical critical exponent, $z$, has
been measured~\cite{Engel,Foot1} to be $z=1$. 

In this paper we report on a numerical study of the localization
properties of single-electron orbitals in double-layer two-dimensional
electron systems.  This work is motivated by recent experiments hinting
at changes in localization properties when two different Landau levels
are nearly
degenerate~\cite{Koch,WeiDeriv,EngelWeiFrac,Engel,WeiSplit,Jiang,Hikami2},
by growing interest in the conditions necessary for the occurrence of
the quantum Hall effect in three-dimensional electron systems, and by
the need for improved understanding of the disappearance of the quantum
Hall effect at weak magnetic fields in high mobility samples.  In each
case, we believe that double-layer quantum Hall systems offer advantages
for both theoretical studies and for the experimental studies which we
hope to motivate.

In single-layer two-dimensional electron systems, localization
properties appear experimentally to be changed when the exchange
enhanced spin-splitting between Landau levels with the same orbital
index collapses.\cite{Fogler} The interpretation of these experiments is
confused by uncertainties involved in modeling the spin-orbit disorder
scattering necessary for mixing the two Landau levels and by the the
apparent importance of interaction effects in controlling the degree of
mixing.  The interaction complications are not so troublesome in
double-layer systems and, in addition, the degree of mixing between
Landau levels in separate quantum wells can be controlled by adjusting
the strength of the barrier separating the wells or by adding an
external bias potential which moves the double-layer system off balance.

In high-mobility two-dimensional electron systems ($E_F \tau / \hbar >>
1$), the quantum Hall effect appears to become unobservable in practice
once Landau level mixing by disorder becomes strong, {\it i.e.} once
$\omega_c \tau $ is of order one.  (Here $\omega_c = e B/ m^{*} c $ is
the cyclotron frequency.)  The loss of an observable quantum Hall effect
in these systems appears to be associated with a dramatic increase in
the localization length in the middle of the Hall plateaus, rather than
with the `floatation' of extended state energies
~\cite{Khmelnitskii,Laughlin,Kravchenko,Glozman,Shahbazyan,YangBhatt,Liu}
which occurs in more strongly disordered systems.  It seems likely that
the same dramatic increase in localization lengths on Hall plateaus will
occur in double-layer systems when the Landau levels in the two-layers
are strongly mixed.   The ability to systematically control the number
of Landau levels which are mixed motivates working with double and
multi-layer systems.  

Since much of the physical picture underlying the quantum Hall effect is
specific to two dimensions, it was not initially clear that it was even
possible to observe quantized plateaus in three-dimensional systems.
Early experimental work, focused on widely separated 2D
layers~\cite{Haavasoja}, found the IQHE with quantized resistivity
$\rho_{xy}=h/N_zie^2$ where $N_z$ is the number of quantum wells,
consistent with parallel conduction in many quantum wells.  St\"ormer
{\it et al.}~\cite{Stormer} performed experiments on coupled GaAs
superlattices with 30 periods, where the dispersion relations explicitly
showed three-dimensional effects in zero magnetic field.  Measurements
of the resistivity tensor showed that $\rho_{xy}$ again was quantized as
$h/nie^2$ despite the 3D nature of the system, but it was found that
$n\neq N_z$.  This was later explained in terms of band bending which
raised the energy of states in quantum wells close to the sample surface
above the Fermi level.~\cite{Ulloa}.  Subsequent experiments on a
200-period superlattice seem to confirm this picture since only a fixed
number of apparently empty quantum wells occur~\cite{Hoffman}.  In later
work St\"ormer {\it et al.}~\cite{Storm2} demonstrated that
$\sigma_{zz}$ also has deep minima on quantum Hall plateaus.
Significantly for the physics addressed here, no Hall plateaus with
intermediate integer indices were observed. 

The phase diagram of disordered three-dimensional systems in a magnetic
field has been investigated~\cite{Murzin,Laue,Zeitler,Wang} and the
possibility of a metal-insulator (MI) transition has been pursued.  The
study of the quantum Hall effect in three-dimensional systems may also
be relevant to quasi-one-dimensional systems such as the Bechg\aa rd
salts which form a spin density wave (SDW) state in a magnetic
field~\cite{Gorkov}.  In such a state the Hall conductance is also
quantized. Recent theoretical~\cite{Balents,BechgaardTheo} as well as
experimental work~\cite{BechgaardExp} supports a picture in which a
complex phase diagram arises from the interplay between SDW formation
and the quantum Hall effect.

Theoretical model calculations have examined the Anderson transition in
a strong magnetic field for three-dimensional systems.  Transfer matrix
calculations~\cite{Ohtsuki1,Henneke} and recursive Green's functions
studies~\cite{Ohtsuki2} have shown that the localization length at the
mobility edge diverges with an exponent $\nu=1.35\pm0.15$~\cite{Henneke}
similar to what is found in the absence of a magnetic
field~\cite{Kramer}.   Calculations performed on network
models~\cite{ChalkDohm} indicate an exponent of $\nu=1.45\pm 0.25$ in,
perhaps, surprisingly good agreement. Note that the exponent $\nu$ is
significantly smaller than the approximately $7/3$ found at the $2D$
quantum Hall transitions.

\begin{figure}
\centering
\epsfxsize=7.5 cm
\leavevmode
\epsffile{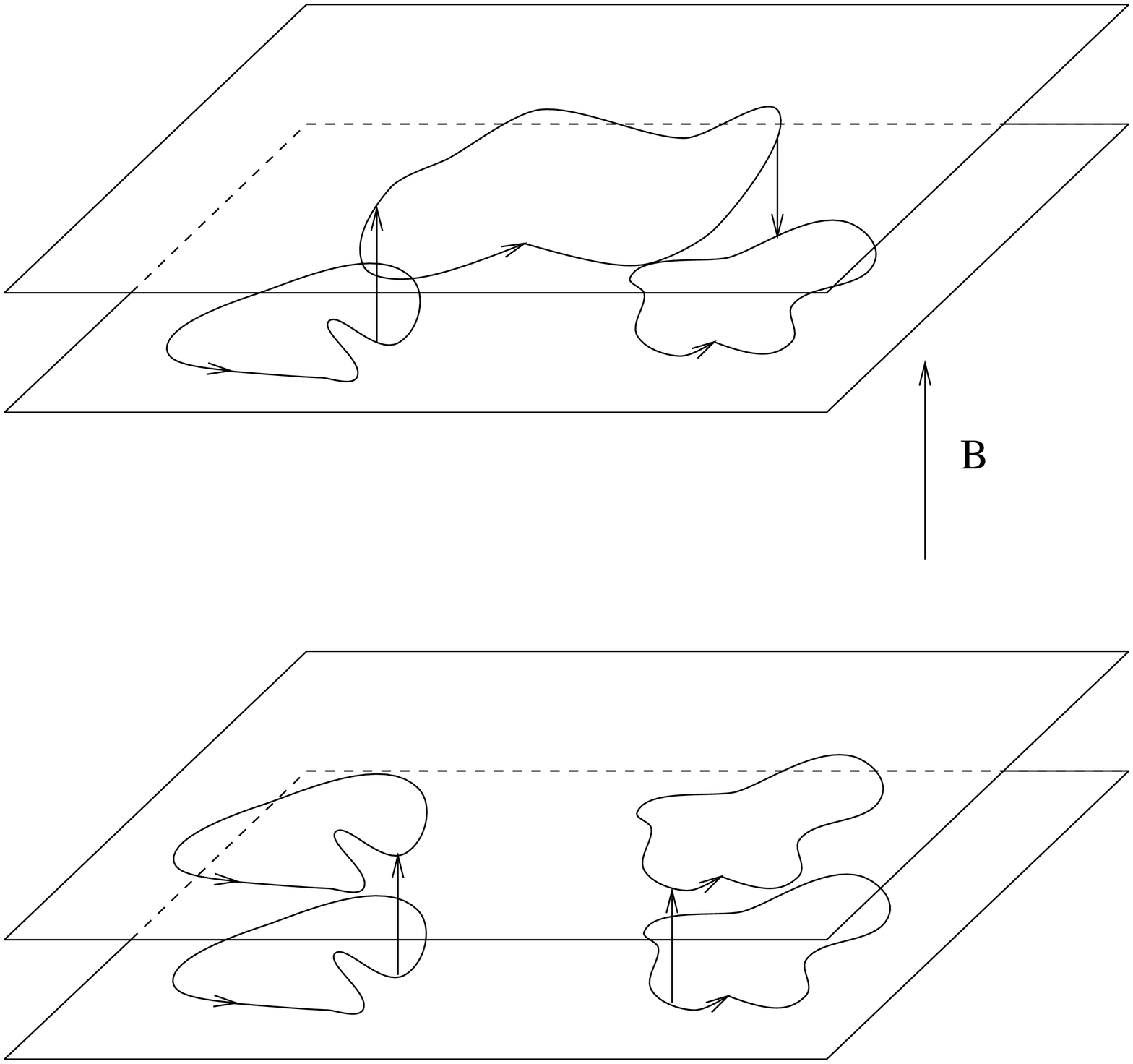}
\caption{The two different forms of disorder.}
\label{fig:disorder}
\end{figure}

In our studies of double-layer systems we assume that the disorder
potentials in the two layers (see Fig.~\ref{fig:disorder}) are
uncorrelated.  As sketched in Fig.~\ref{fig:disorder} it is crucial to
make the distinction between this form of disorder, which we shall refer
to as uncorrelated disorder, and the form where the disorder is
identical in the layers (correlated disorder) since the latter has a
much smaller effect on the localization length.   Related work on the
double-layer system has previously been done by Ohtsuki {\it et
al.}~\cite{Ohtsuki}.   Our study has clear analogies to previous work on
the spin-degenerate (where the two spin levels are not resolved) and
spin-resolved transitions.  Experimentally, the spin-degenerate
transition has been investigated in several different studies.  Wei et.
al.~\cite{WeiDeriv} observed that $\Delta B$ as well as
$(d\rho_{xy}/dB)^{-1}$ both behave as a power-law in $T$ with a much
smaller exponent than observed for the spin-resolved transitions. Later
experiments~\cite{WeiSplit} found $\Delta B\sim T^{0.21}$. Experiments
on the frequency, $(f)$, dependent conductivity have shown that
$\sigma_{xx}$ peaks broadens as $(\Delta B)\sim f^\gamma$ with
$\gamma=0.41\pm0.04$ for the spin split transition and $\gamma=0.20\pm
0.05$ for the spin-degenerate case~\cite{Engel}. This, tantalizing
effect was explained in terms of an unstable critical point so that the
enhancement of the exponent is understood as an artifact of crossover
phenomena~\cite{WeiSplit} (See Fig.~\ref{fig:flow}).  However, other
work~\cite{Fogler} has proposed the possibility of an effectively {\it
stable} fixed point due to a disorder induced destruction of the
exchange enhancement of the electron $g$-factor.  In the double-layer
system this would correspond to a finite critical coupling $t_c$ needed
to see the symmetric and antisymmetric state appear.  Network model
calculations~\cite{Derek,WLW} find that the universality class is
unchanged in the presence of strong Landau level mixing between the
polarized Landau sub-bands. However, it is presently not clear to what
extent these models actually describe Landau level
mixing~\cite{ChalkDohm,Shahbazyan}. The effect of spin-orbit scattering
has also been considered~\cite{Charles,Polyakov2} again leading to crossover
phenomena without a change of the universality class.  In the language
of the double-layer systems these studies consider completely random
tunneling between the two layers and in most cases the disorder is
strongly correlated between the two layers.  Since the experimental
situation for double-layer systems is much closer to a {\it weak} and
{\it uniform} tunneling between the two layers, which was not considered
previously, we investigate this limit in detail. In the case of the
spin-degenerate spin-resolved transition the controlling parameter is
the electron $g$-factor which is not readily tuned experimentally.
However, for double-layer systems, it should be experimentally possible
to tune the coupling between the two layers by tilting the magnetic
field. This is a very convenient circumstance since it is (conceivably)
possible to investigate if a finite coupling $t_c$ is needed in order to
split the Landau level subbands corresponding to the symmetric and
antisymmetric state.  Assuming that the interesting interaction effects
associated with tilting the magnetic field~\cite{Yang,MoonI,MoonII} can
be ignored in the more disordered samples to which the considerations of
the present paper would apply, the only effect of tilting the magnetic
field in a sample with uncorrelated disorder is to reduce~\cite{JunHu}
the effective tunneling parameter:
\begin{equation}
t^{*} = t \exp ( -  (d \tan (\theta)/2 \ell)^2) 
\label{eq:ttilt}
\end{equation}
where $d$ is the layer separation, $2 \pi \ell^2 B_{\perp} = \Phi_0$
with $B_{\perp}$ the component of the magnetic field perpendicular to
the sample, and $\theta$ the angle by which the magnetic field is tilted
away from the normal to the layers.  

Our paper is organized as follows.  In Section~\ref{sec:args} we discuss
several different arguments, all at the semiclassical level, which
indicate that even very weak coupling between the layers can give rise
to dramatic increases in localization lengths and possibly change the
apparent localization length exponent.    In Section~\ref{sec:diags} we
present the results of a numerical study for a realistic microscopic
model of a double-layer system.  Localization properties are discussed
in terms of the Thouless number $g_s$. Section~\ref{sec:network} is
concerned with network model suitably defined to describe the weakly
coupled double-layer system. Using transfer matrix techniques we
calculate the reduced correlation length.   Finally in
Section~\ref{sec:conclusion} we present our conclusions.

\section{Semiclassical Arguments}\label{sec:args}

When the disorder potentials in both layers are smooth on 
microscopic length scales electronic orbitals are localized 
along equipotentials.  The quantum percolation theory of 
the integer quantum Hall effect is based on a theory of 
percolating equipotentials supplemented by the possibility
of quantum tunneling between equipotentials near saddle points 
of the disorder potential.   Semiclassically, the `E cross B' drift 
velocity of an electron along an equipotential is proportional
to the local electric field.  For a Landau level of width 
$\Gamma$ the typical electric field is 
$ \Gamma /e \Lambda$ where $\Lambda$ is the correlation
length of the disorder potential.  Accordingly the typical
drift velocity is 
\begin{equation}
v_{dr} = \frac{c \Gamma}{e B \Lambda}.
\label{eq:driftvel}
\end{equation}
We are interested in the influence of tunneling between 
the layers on localization properties.  Tunneling introduces 
a new typical length scale into the physics of the system,
the drift length
\begin{equation}
l_{dr} = \frac{\ell^2 \Gamma}{\Lambda t}.
\label{eq:driftlength}
\end{equation}
$l_{dr}$ is the typical distance an electron drifts along an
equipotential between tunneling events.  Typical equipotentials 
are closed paths with a perimeter which we can for 
present purposes associate with the localization length $\xi$ which 
diverges at a critical energy within each Landau
level~\cite{Luryi,Trugman}.  
For weak tunneling $l_{dr}$ is long.  When $l_{dr}$ is larger
than the sample size, tunneling will have no effect on 
even the most extended orbitals in the system and we should expect
negligible changes due to tunneling in our finite size 
system calculations.  When $l_{dr}$ is smaller than $\xi$,
electrons will typically tunnel before completing a closed
orbit around an equipotential.  Note that this 
condition is satisfied at smaller and smaller $t$ values as
the center of the Landau level is approached.  When 
an electron tunnels to the other layer, it will move
along an equipotential of a statistically independent smooth
disorder potential.   When it later returns to the original layer,
it will not in general return to the same equipotential contour
from which it departed {\it even neglecting tunneling events near
saddle points within a layer}. Related ideas have been discussed
in Ref.~\onlinecite{Polyakov2}. In our view the possibility that
localization physics is qualitatively altered by tunneling
between the layers deserves serious attention.  The 
situation becomes simple again only when $l_{dr}$ is smaller
than $\Lambda$ so that an electron tunnels many times before
its local potential profile changes.  In this limit,
electronic eigenstates will be symmetric and 
antisymmetric combinations of the individual layer eigenstates
and the effective
disorder potential will be the mean of the independent 
disorder potentials in the two layers.  Note that the limit
of infinite system size and the limit of vanishing tunneling
amplitudes are not interchangeable.  In the thermodynamic
limit, tunneling will always be important for those states
near the critical energy which have a localization length
larger than $l_{dr}$.  

\begin{figure}
\centering
\epsfxsize=7.5 cm
\leavevmode
\epsffile{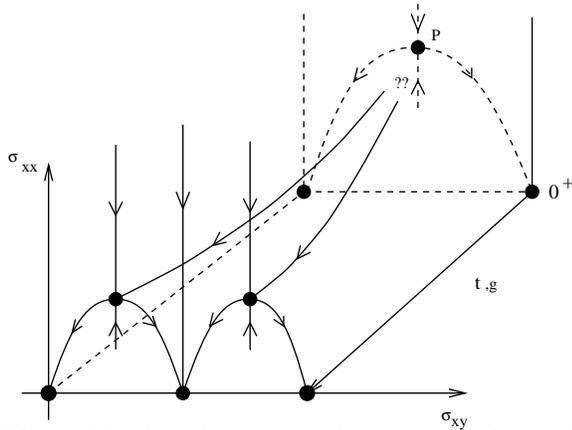}
\caption{The flow diagram in the presence of tunneling, from
Ref.~\protect\onlinecite{WeiSplit}.}
\label{fig:flow}
\end{figure}

In Fig.~\ref{fig:flow} we show the flow diagram proposed in
Ref.~\onlinecite{WeiSplit}. The controlling parameter that
describes the flow from the unstable fixed point, $P$, to the
more conventional picture (solid lines) is the electron $g$-factor
or the tunneling parameter, $t$, for the double layer systems.
Note that the unstable fixed point is at $0^+$ since two uncoupled
layers ($t=0$) cannot lead to new critical behavior.
In Fig.~\ref{fig:flow} we have left out the flow around the
unstable critical point $P$ leaving open the possibility of a
new unstable critical point at a finite $t_c$. Such an unstable
critical point would imply that the transition from the
$\nu=0$ phase would be directly into $\nu=2$ phase without an
intervening $\nu=1$ state. This scenario has recently been discussed
by Tikofsky and Kivelson~\cite{Tikofsky} 
and recent experiments in the strong
disorder weak field limit could be interpreted
as lending support to this possibility~\cite{Jiang,Wong,Wang2,Hughes}.

If, as suggested in Ref.~\onlinecite{WeiSplit}, the enhanced exponent
seen in the experiments on the spin-degenerate transition is indeed
due to the presence of an unstable fixed point one might ask why
the crossover exponent 
should be $14/3$. Polyakov {\it et al.}~\cite{Polyakov}
have proposed a crossover form for the correlation length based on the
assumption that the effective correlation length is the square of
the correlation length in the absence of any coupling. We now give
a brief argument, somewhat 
speculative and heuristic, in the spirit of the semiclassical calculation
of Ref.~\onlinecite{MilSok}, for why this could be the case. In
the absence of any tunneling to the other layer we will at a given
energy $E$, relative to the middle of the Landau level, have states
localized on a equi-potential contour $\xi_0(E)$. We then introduce 
a very weak coupling, $t$, to the other layer. Following the
discussion in the beginning of this section we expect that
for $\xi_0(E)\ll l_{dr}$
there will not be sufficient time to scatter into the next layer
before the electron self-interacts. Thus a non-zero $t$ will not
affect sufficiently small $\xi_0(E)$. However, when $\xi_0(E)\sim l_{dr}$,
but still in the limit $t\ll \Gamma$ the electron will scatter
a number of times, $N_{\rm s}$, along the {\it final}
path $\xi(E)$. Naturally,
\begin{equation}
N_{\rm s} \sim \xi_0(E)/l_{dr}.
\end{equation}
Furthermore, in the limit $t\ll \Gamma$ we expect the electron to 
be scattered among {\it different} orbits in {\it different layers}
at approximately the 
{\it same} energy and spatial extent, $\xi_0(E)$ (see top panel in
Fig.~\ref{fig:disorder}). We then
find that
\begin{equation}
\xi(E)\sim N_s(E)\xi_0(E) \sim (\xi_0(E))^2\sim|E|^{-2\nu}.
\end{equation}
In a given orbit of extent $\xi_0(E)$, which makes up a part of
the final path of length, $\xi$ this argument assumes only a few scattering
events into {\it different} orbits. That is, we implicitly exclude
events where the electron immediately is scattered back into the same orbit.
Clearly when $t$ becomes
large many scattering events will occur and we effectively form
symmetric and antisymmetric states. This will occur when
$t\gg \Gamma$ or equivalently
\begin{equation}
l_{dr}\ll \Lambda.
\end{equation}
Thus we expect an enhanced correlation length and an effective
doubling of the correlation length
exponent whenever
\begin{equation}
\Lambda\ll l_{dr}\ll \xi.
\end{equation}

\section{Exact Diagonalization Results}\label{sec:diags}

We now turn to a discussion of our exact diagonalization results
for double-layer systems. 
We shall work exclusively in the lowest Landau level approximation.
Since we are considering a finite system of dimensions 
$L_x, L_y$ we want to impose periodic boundary conditions.
In this case one uses the following set of basis functions
for the lowest Landau level~\cite{Ando,AOL}:
\begin{equation}
\phi_j^n(x,y)=\sum_{s=-\infty}^{\infty}
\left({\frac{1}{L_y\ell\sqrt{\pi}}}\right)^{1/2}
e^{i\frac{X_{j,s}y}{\ell^2}}
e^{-\frac{\left(x-X_{j,s}\right)^2}{2\ell^2}}.
\label{eq:wfctn}
\end{equation}
Here, $X_{j,s}=j2\pi \ell^2/L_y+sL_x$ and $\ell=\sqrt{\hbar c/2eB}$
is the magnetic length
and $j$ runs from $1$ to $N_\phi = L_xL_y/2\pi \ell^2$ where
$N_\phi$ is the number of flux quanta or the degeneracy
factor of the lowest Landau level. Since we describe
each quantum well by this set of lowest Landau level
wave-functions we include the dummy index $n$ to denote the
different layers.
It is easy to see
that the individual terms in the infinite sum all are 
invariant under $y\to y+L_y$. The sum over $s$ makes the
wave-function invariant up to phase factors also under the transformation
$x\to x+L_x$.
We model the randomness as $\delta$-function scatterers
at random position with random sign. 
\begin{equation}
V({\bf r}) = 2\pi\lambda \ell^2 \sum_{p=1}^{n_p}\delta({\bf r}-{\bf r }_p),
\end{equation}
Although this is
not a realistic model of real randomness it is generally believed
that the form of the randomness is irrelevant, see however
Ref.~\onlinecite{LiuSarma}. 
The Hamiltonian
can then be written 
\begin{eqnarray}
<\phi_i^n|&H&|\phi_j^{n'}>=\nonumber\\
& &2\pi\lambda \ell^2\delta_{n,n'}
\sum_{p=1}^{n_p}
\phi^{*n}_i({\bf r}_p)\phi_j^{n'}({\bf r}_p)\nonumber\\
& &-t\delta_{i,j}(\delta_{n',n-1}+\delta_{n',n+1}),
\label{eq:H}
\end{eqnarray}
where $\lambda$ contains the random sign of the $\delta$-function
scatterers and their strength. $n_p$ is the number of scatterers.
The first term in Eq.~(\ref{eq:H}) is the potential energy in each of
the wells while the second describes the tunneling between the two wells
in the tight-binding approximation, with $t$ the tunneling parameter.
We shall always use periodic boundary conditions in the $z$-direction
so that layer $N_z+1$ is identical to the layer 1.
We chose $|\lambda|$ so as to fix the width
of the Landau levels to be of the order of 1 in the self-consistent
Born approximation~\cite{AOL}, $\Gamma_{\rm SCBA}$, in our units we have
\begin{equation}
\Gamma_{\rm SCBA}=2|\lambda|\sqrt{n_p/N_\phi},
\end{equation}
and we therefore chose $|\lambda|=(1/2)\sqrt{N_\phi/n_p}$.
With this choice we have effectively chosen our energy scale
and all of our results have energy, $E$, in units of
$\Gamma_{\rm SCBA}$.
For different sizes we always keep $|\lambda|$ and therefore
the ratio $n_p/N_\phi$ constant. From exactly solvable
models~\cite{Wegner} it is known that the density of states
exhibit peculiarities when $n_p/N_\phi\le 2$ we have
therefore chosen always to work with $n_p/N_\phi=3$.
We have explicitly checked that in the case of a single layer
we find, with the above mentioned definitions that the calculated density
of states is well described by the following approximate formula
\begin{equation}
2\pi\rho(E)=\sqrt{\frac{2}{\pi}}\frac{\hbar}{\Gamma}
e^{-2\left(\frac{E}{\Gamma}\right)^2},
\label{eq:rSCBA}
\end{equation}
with, $\Gamma=1$. In the case where we do not include the random sign
of the scatterers we have also checked that the density of states
corresponds to the exact result of Ref.~\onlinecite{Wegner}.
Note that we use the standard definition of $\rho(E)$ which
integrates over $E$ to $1/(2\pi l^2)$.

We note that with our definition of the potential
the variance of $V$:
\begin{equation}
<V({\bf r})V({\bf r'})>=6\pi \ell^2\lambda^2\delta({\bf r-r'})\equiv
v^2\ell^2\delta({\bf r-r'}).
\end{equation}
With our choice of $\lambda$ we then have $v=\sqrt{\pi/2}$.
We could equally well have chosen our energy scale by making
the choice $v=1$, and thereby fixing $\lambda$.

\subsection{Computational Method}

The Thouless number~\cite{Thouless}, $g_s(E)$,
is defined as the absolute value of the shift of a given
energy level, under a change in boundary conditions from periodic
to antiperiodic, $|\Delta E|$, multiplied by the total density
of states, $N(E)$,
\begin{equation}
g_s(E)=N(E)\Delta E.
\end{equation}
Here $N(E)$ integrated over all energies is the total number
of states, $N_zN_\phi$, for an $N_z-$layer system, or
in other words, $N(E)=L^2N_z\rho(E)$. Clearly extended
states are much more sensitive to a change in the boundary conditions
than localized, and $g_s(E)$ therefore measures the stiffness at a
given energy, $E$. In our calculations we change the boundary conditions
in the y-direction
in the individual layers simply by performing the transformation
$X_{j,s}\to X_{j+1/2,s}$ in Eq.~(\ref{eq:H}).
It is known that in the absence of a magnetic field
the Thouless number is related to the longitudinal conductivity,
$\sigma_{xx}(E)$~\cite{Edwards}. For a recent account of Thouless
number calculations see Ref.~\onlinecite{Charles} and references therein.
We follow Ref.~\onlinecite{Charles} in deriving a scaling function
for the integrated Thouless number.
If we assume that the correlation length diverges as $\xi\sim|E|^{-\nu}$
we can write a finite-size scaling form for $g_s(E)$:
\begin{equation}
g_s=\tilde g_s(E L^{1/\nu}),
\end{equation}
where $\tilde g_s$ is a universal function. From this it follows that
\begin{eqnarray}
A(L)&=&\int_{-\infty}^{\infty}g_s(E)dE\nonumber\\
&=& L^{-1/\nu}\int_{-\infty}^{\infty}\tilde g_s(x)dx=C L^{-1/\nu},
\label{eq:AL}
\end{eqnarray}
where $C$ is a constant independent of the system size.
In order to perform the disorder averaging we consider 
40,000 samples for $N_\phi=12$, 10,000 for $N_\phi=20$,
2,000 for $N_\phi=80$, 1,000 for 
$N_\phi=300$, and 200 for $N_\phi=1,000$.
In each case the system consists
of $N_z=2$ layers each with $N_\phi$ states. We have also
preliminary results for $N_z>2$.
Since we exclusively consider
systems with $L_x=L_y$ we shall in the following use
$L=\ell\sqrt{2\pi N_\phi}$. 

\begin{figure}
\centering
\epsfxsize=8.5 cm
\leavevmode
\epsffile{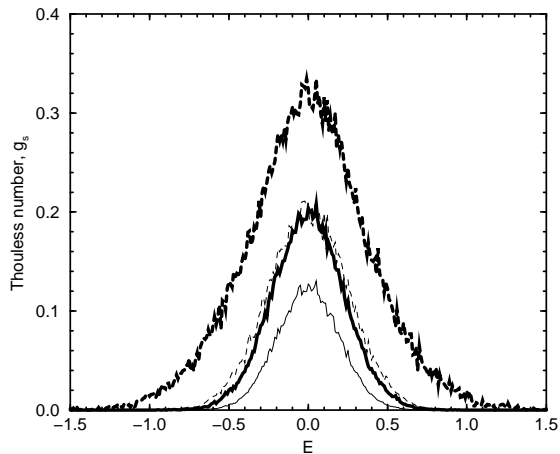}
\caption{The Thouless number, $g_s$, as a function of energy
for a single layer system with $N_\phi=72$, and 
$n_p=5N_\phi$. Calculations
are shown using the elliptic theta functions (thick solid line)
as well as without them (thick dashed line).
We also show results for $g_s$ calculated using a geometric mean
and the theta-functions (thin solid line) and without them
(thin dashed line).
}
\label{fig:comp}
\end{figure}

Building on the work of Thouless and coworkers~\cite{Thouless,Edwards}
that relates $g_s(0)$ to $\sigma_{xx}$ Ando~\cite{Ando} has proposed
that a similar relation should hold in a magnetic field,
\begin{equation}
\sigma_{xx}(E)=\lim_{L\to\infty}\frac{\pi}{2}\frac{e^2}{h}g_s(E,L).
\end{equation}
This is however not true in any strict sense and one would
only expect the left hand side of the above expression to be
proportional to $\sigma_{xx}$.
However, in the absence of a 
magnetic field the root-mean-square level curvature can be related
to the dissipative conductance~\cite{Akkermans}.
If the scaling theory~\cite{Pruisken} of the quantum Hall effect is correct
we would expect that $\sigma_{xx}=(1/2)e^2/h$ at the critical point
and thus $g_s(0)=1/\pi$. This seems to be consistent with what is 
found numerically for short-range scatterers~\cite{Ando,Charles}
although the range of the potential can change the value
significantly~\cite{Ando2,Charles,LiuSarma}. A more rigorous approach
would be to calculate the Chern
numbers~\cite{Niu,Chern,HuoBhatt,HuoHetzelBhatt,YangBhatt}, 
which confirms the
results of the scaling theory.
In our calculations we find $g_s(0)\sim 0.2$, (see Fig.~\ref{fig:comp})
where calculations are shown with (thick solid line) or without
(thick dashed line)
the sum over $s$ in Eq.~(\ref{eq:wfctn}). Clearly the results are
markedly different. Also the density of states differ significantly
without the infinite sum in Eq.~(\ref{eq:wfctn}) we find a density
of states that is no longer well described by Eq.~(\ref{eq:rSCBA}).
In Fig.~\ref{fig:comp} we also show results where the geometric mean
$\exp(<\ln|\Delta E|>_{\rm av})$ has been used instead of $|\Delta E|$. 
With the sum over $s$ in Eq.~(\ref{eq:wfctn}) the Thouless number is
indicated as the thin solid line in Fig.~\ref{fig:comp} and without
the sum as the thin dashed line.

\subsection{Results}
Before we turn to a discussion of the numerical results 
let us begin by looking at a few simple limits:

{\it Correlated disorder:} In this case we can treat
the $N_z-$layer case straight forwardly. By Fourier-transforming
along the $z-$direction we remove the off-diagonal tunneling
elements and obtain a matrix that is block diagonal with
$N_\phi\times N_\phi$ blocks, ${\bf M}_n$. If we denote the wave-vector
along the z-direction by $k_n^z=2\pi n/L_z, \ n=1,\ldots N_z$,
where $N_z=L_z/a$ is the number of layers,
we find that each of these blocks can be written:
\begin{equation}
{\bf M}_n={\bf B}-2t\cos k_n^z{\bf I},
\end{equation}
where ${\bf B}$ is the matrix describing the disorder in 
one layer. The presence of the $2t\cos k_n^z{\bf I}$ term will
not affect the Thouless numbers since it is independent of the
boundary conditions in the $xy$ plane. Thus, except for some
accidental degeneracies we should find that the Thouless numbers
are a simple superposition of the the 1-layer result displaced  
by $2t\cos k_n^z$:
\begin{equation}
g_s(E)=\sum_n g_s^{\rm 1-layer}(E-2t\cos k_n^z).
\label{eq:gscol}
\end{equation}

{\it Strong tunneling, uncorrelated disorder:} Now we consider the
case of uncorrelated disorder. In the limit where $t$ tends to $\infty$
the tunneling completely dominates over the disorder and it is
again advantageous to perform a Fourier transform in the $z-$direction.
In the limit $t\to\infty$ we can neglect the block-off-diagonal
matrices and we again obtain a block-diagonal matrix with
$N_\phi\times N_\phi$ blocks, ${\bf M}_n$. However, this time
we find:
\begin{equation}
{\bf M}_n=\frac{1}{N_z}\sum_m{\bf B}_m-2t\cos k_n^z{\bf I},
\end{equation}
where ${\bf B}_m$ is the matrix describing the disorder in 
the $m$th layer. Let us consider the two-layer case, $N_z=2$. 
For $\rho(E)$ we obtain
two widely separated peaks each corresponding to a single
layer with double the number of impurities at half the strength.
Each of these peaks then have a width $\widetilde\Gamma=\Gamma/\sqrt{2}$.
Hence, the integrated Thouless number, $A(L)$, should {\it increase}
by a factor of $\sqrt{2}$ compared to the single layer result.

{\it Zero tunneling, uncorrelated disorder:} If we set  $t=0$ the 
Hamiltonian matrix again becomes block-diagonal and we should
obtain results similar to the single layer case for large
enough system sizes. The density of states, $\rho(E)$, should remain
unchanged and $g_s(E) \sim N_zg_s^{\rm 1-layer}(E)$, since the
total energy of states, $N(E)$, is proportional to $N_z$.

\begin{figure}
\centering
\epsfxsize=7.0 cm
\leavevmode
\epsffile{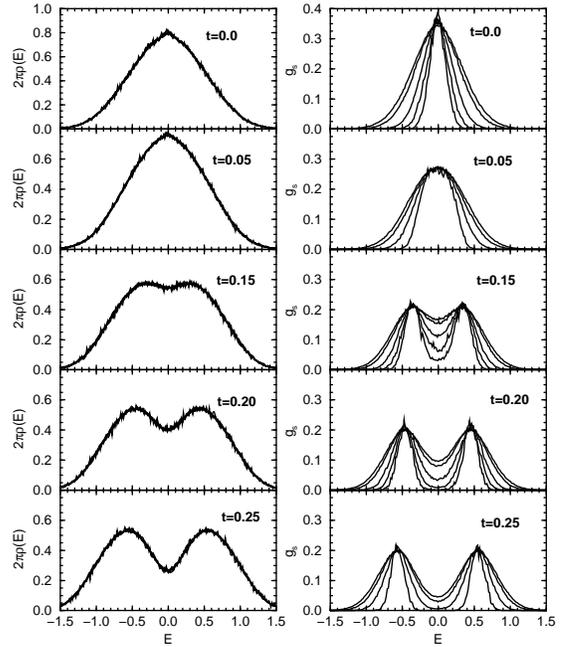}
\caption{The density of states, $\rho (E)$, and the Thouless number,
$g_s$, for a range of couplings $t$ between the two layers. 
$n_p=3N_{\phi}$ random pinning centers, modeled as random
sign $\delta$-functions, were used in the calculation. The position
of the pinning centers was taken to be random and independent in the
two layers, corresponding to uncorrelated disorder. 
In all figures the curves corresponds to $N_\phi=12,20,80,300,1000$.
The energy $E$ is in the units of $\Gamma_{\rm SCBA}$.
The dashed lines indicates the position 
of the extended state energies in the
absence of disorder.
}
\label{fig:bigfig}
\end{figure}

We shall mainly be concerned with disorder that is not correlated
between the layers (the top panel in Fig.~\ref{fig:disorder})
but we shall briefly also discuss the case of correlated disorder
(the bottom panel in Fig.~\ref{fig:disorder}).
The bulk of our results are shown in Fig.~\ref{fig:bigfig} where we
display the density of states, $\rho(E)$,
along with the Thouless number, $g_s$, for a range of couplings,
$t=0.0,\ 0.05,\ 0.15,\ 0.20,\ 0.25$, between the two layers.
In all cases the disorder was taken to be independent in the two layers.
For the uncorrelated disorder model we consider here, 
the effective correlation
length of the disorder potential is the microscopic length $\ell$.  We 
therefore expect that tunneling will be important when the 
localization length of decoupled layers 
exceeds $\ell_{dr} \sim \ell \Gamma/t$,
or, equivalently, $\xi$ exceeds $l_{dr}$. 
Accordingly the influence of tunneling on the 
density of states and especially on the Thouless numbers appears
first near the center of the levels where the localization lengths
are large.  Well separated localized states appear 
for sufficiently large $t / \Gamma $.  
In the top row our results for two uncoupled layers, $t=0.0$, are shown.
We see that the Thouless number is exactly twice the value of a single
layer (see Fig.~\ref{fig:comp}). For very weak tunneling the peak
in $g_s$ becomes significantly broadened and for the sizes we have
considered we do not observe two separate peaks until $t\sim0.1$.
As the tunneling between the two layers, $t$, is increased two
peaks corresponding to the antisymmetric and symmetric state become
apparent. For $t=0.25$ we find that $g_s$ and the density of
states essentially behave as the superposition of two independent peaks
for the symmetric and antisymmetric state in agreement with our
considerations above. We also observe that the width of the density
of states in the individual peaks approximately obeys the
relation $\Gamma\sim\Gamma_{\rm 1-layer}/\sqrt{2}$. 
Our results are in good agreement with prior
calculations by Ohtsuki et. al~\cite{Ohtsuki}.
In fig.~\ref{fig:bigfig} the position of the degenerate Landau levels in 
the absence of disorder is indicated as dashed lines in the panels
for $g_s$. For large $t$, level repulsion is clearly visible and the
two extended state energies are further apart when disorder is included.

\begin{figure}
\centering
\epsfxsize=8.5 cm
\leavevmode
\epsffile{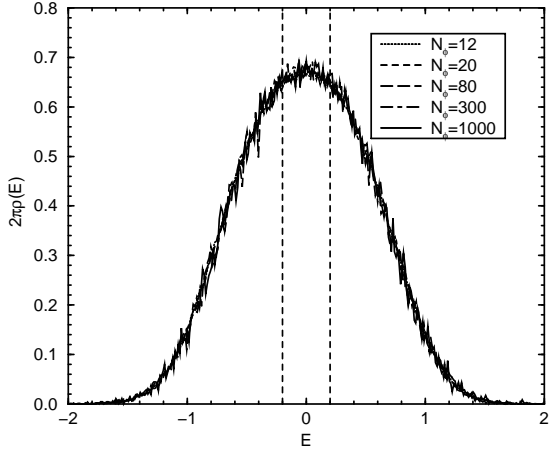}
\caption{The density of states for a coupling $t=0.1$ between the two
layers. $n_p=3N_{\phi}$ random pinning centers, modeled as random
sign $\delta$-functions, were used in the calculation. The position
of the pinning centers was taken to be random and independent in the
two layers. The two dashed lines indicates the position of the two
degenerate Landau levels in the absence of any disorder.}
\label{fig:gb0.1}
\end{figure}

\begin{figure}
\centering
\epsfxsize=8.5 cm
\leavevmode
\epsffile{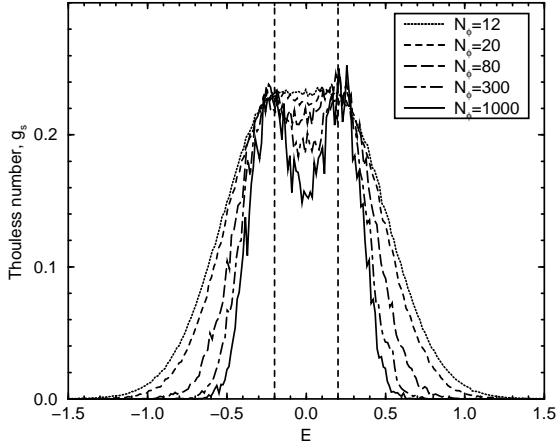}
\caption{The Thouless number, $g_s$, for a coupling $t=0.1$ between the two
layers. $n_p=3N_{\phi}$ random pinning centers, modeled as random
sign $\delta$-functions, were used in the calculation. The position
of the pinning centers was taken to be random and independent in the
two layers. The two dashed lines indicates the position of the two
states in the absence of any disorder. The results corresponds
to the density of states in Fig.~\protect\ref{fig:gb0.1}}
\label{fig:gs0.1}
\end{figure}

Before discussing the scaling properties of the transition we compare
results for disorder that is independent in the two layers (uncorrelated
disorder) or the same (correlated disorder). In Figs.~\ref{fig:gb0.1},
\ref{fig:gs0.1} we show the density of states and $g_s$ for the case
of uncorrelated disorder in two-layers coupled with a tunneling parameter of
$t=0.1$.
The density of states shown in Fig.~\ref{fig:gb0.1} is markedly broader
than what we found for two uncoupled layers (top row
Fig.~\ref{fig:bigfig}). 
In Fig.~\ref{fig:gs0.1} we show the Thouless numbers for $t=0.1$, only
for the largest sizes does it become clear that in the 
thermodynamic limit extended state energies exist at two discrete 
energies rather than across a band of finite width between low 
and high energy mobility edges.
Note also that the peak value of the
Thouless number is in this case $g_s^{\rm max}(t=0.1)\sim 0.22$ whereas we
found $g_s^{\rm max}\sim 0.20$ for the single layer case and
$g_s^{\rm max}\sim 0.35(t=0.0)$ for two layers in the absence of any
tunneling.

\begin{figure}
\centering
\epsfxsize=8.5 cm
\leavevmode
\epsffile{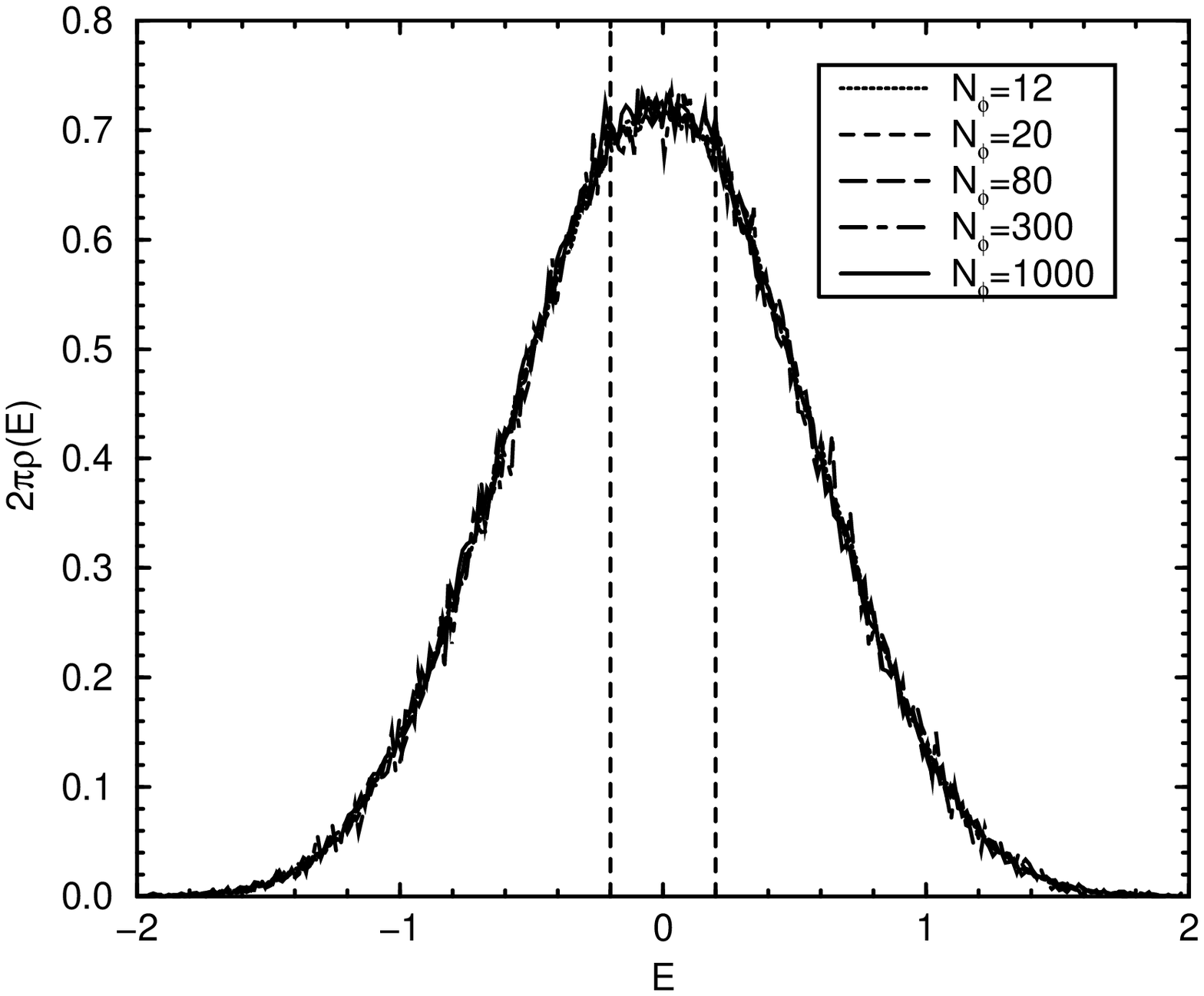}
\caption{The density of states for a coupling $t=0.1$ between the two
layers. $n_p=3N_{\phi}$ random pinning centers, modeled as random
sign $\delta$-functions, were used in the calculation. The position
of the pinning centers was taken to be random but the {\it same} in the
two layers, corresponding to a form of correlated disorder. 
The two dashed lines indicates the position of the two
degenerate Landau levels in the absence of any disorder.}
\label{fig:col.gb0.1}
\end{figure}

\begin{figure}
\centering
\epsfxsize=8.5 cm
\leavevmode
\epsffile{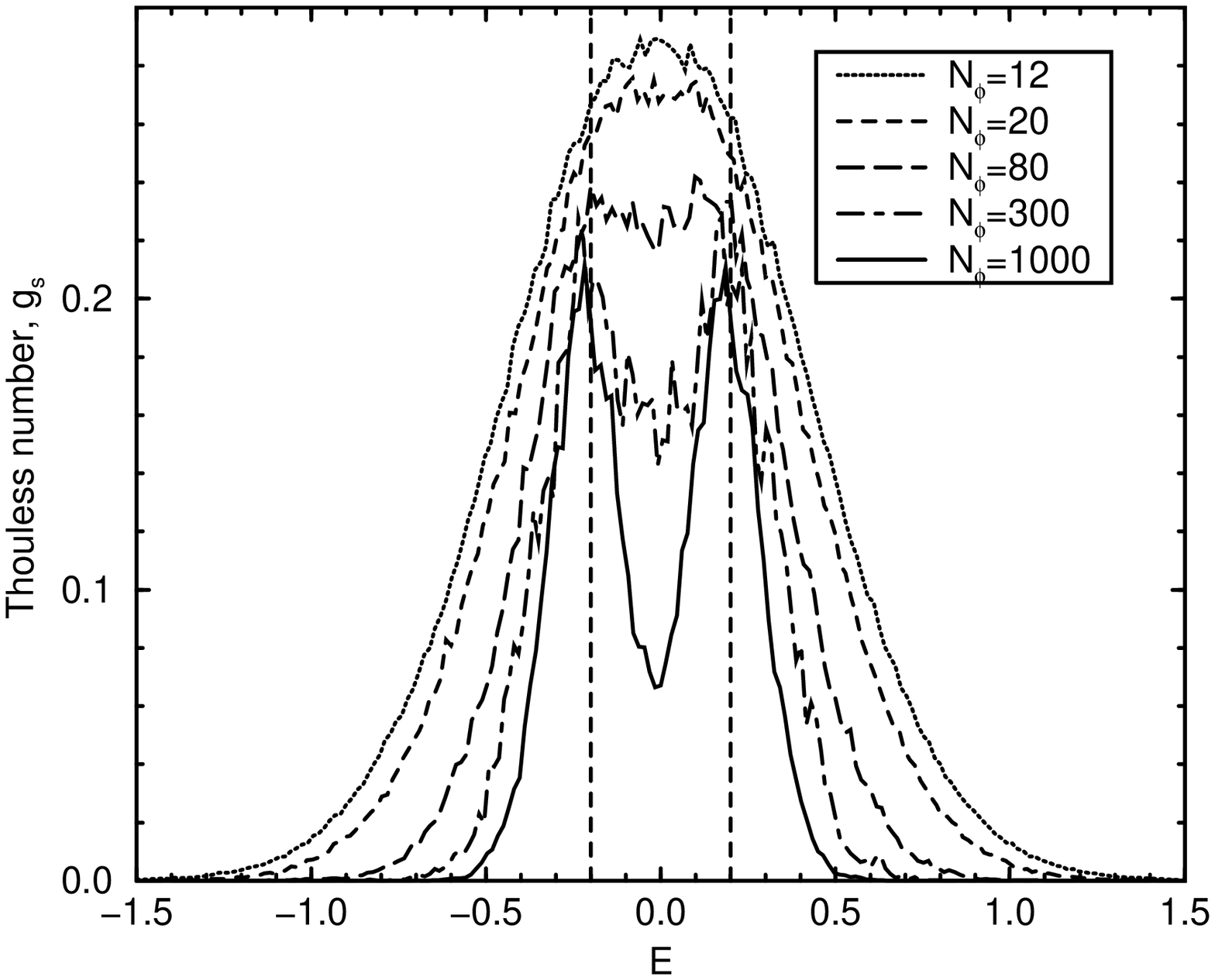}
\caption{The Thouless number, $g_s$, for a coupling $t=0.1$ between the two
layers. $n_p=3N_{\phi}$ random pinning centers, modeled as random
sign $\delta$-functions, were used in the calculation. The position
of the pinning centers was taken to be random and but the {\it same} in
the two layers, corresponding to a form of correlated disorder. 
The two dashed lines indicates the position of the two
degenerate Landau levels in the absence of any disorder. 
The results corresponds
to the density of states in Fig.~\protect\ref{fig:col.gb0.1}}
\label{fig:col.gs0.1}
\end{figure}

The case of correlated disorder is very different. In
Figs.~\ref{fig:col.gb0.1}, \ref{fig:col.gs0.1} we show our results for
the density of states and $g_s$, respectively. The disorder in this
case is the same in the two layers and the tunneling parameter was, as
above, taken to be $t=0.1$.
If we compare $\rho(E)$ for
uncorrelated and correlated disorder to the single 
layer results we see that $\rho(E)$ is decreased at $E=0$ in both cases
but more so for the case of uncorrelated disorder. 
This corresponds to a depletion of
states at the center of the Landau band.
The Thouless number, $g_s$, is also significantly different for the
case of correlated disorder. $g_s^{\rm max}$ is approximately $0.28$
for the smaller sizes before approaching a value of 
$g_s^{\rm max}\sim 0.20$ for the largest systems. As expected the 
Thouless numbers are well described by a simple super position
of single layer results Eq.~(\ref{eq:gscol}). This is clearly
not the case for the results in Fig.~\ref{fig:gs0.1}.

\begin{figure}
\centering
\epsfxsize=8.5 cm
\leavevmode
\epsffile{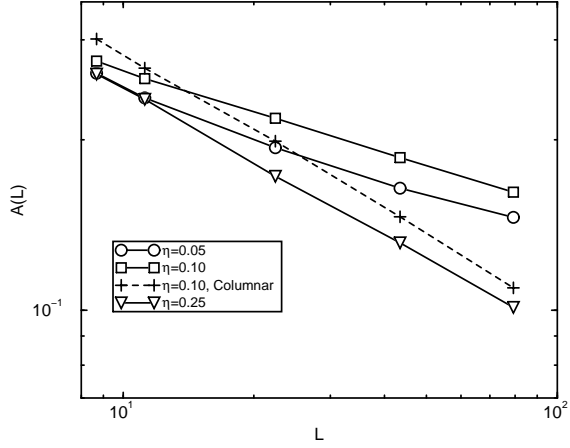}
\caption{The integrated Thouless number as a function of
$L$, for tunneling strenghts of $t=0.05,\ 0.1,\ 0.25$. The
plusses denote results for $t=0.1$ with correlated disorder.}
\label{fig:al}
\end{figure}

The difference between correlated 
and uncorrelated disorder is also reflected
in the scaling of integrated Thouless number, $A(L)$ Eq.~(\ref{eq:AL}).
In Fig.~\ref{fig:al} we show results for 
the integrated Thouless number for three different tunneling
strengths. We see from Eq.~(\ref{eq:AL}) that $A(L)$ according
to the finite-size scaling form should behave as a power-law
in $L$ with exponent $-1/\nu$. For the case of uncorrelated disorder,
the results in Fig.~\ref{fig:al} yield $1/\nu\simeq 0.43$ for
$t=0.25$ in agreement with previous results on single layer systems. 
Correlated disorder (plusses in Fig.~\ref{fig:al}) behaves in a similar
way and we find $1/\nu\simeq 0.44$.
However, the case of uncorrelated disorder with weak tunneling shows
a marked crossover. For $t=0.1$ we find $1/\nu\simeq 0.24$.
(Interestingly, numerical calculations for single-layer systems
in the $N=1$ Landau level show similar apparent enhancement of the
localization length exponent.)
For $t=0.05$ we see that the smaller system sizes 
show the $1/\nu \simeq 0.44$ behavior expected for decoupled 
systems before crossing over to a different power law.
For small $t$ and short-range potential correlations 
we do not expect tunneling to have any effect
until the system size reaches $\sim 1/t$.  
If we fit only to points
with $L>15$ we find, for $t=0.05$, $1/\nu\simeq 0.23$ 
in very good agreement with the result for $t=0.1$.
For $t=0.15$ and $t=0.2$ we find in both cases that the slope
of $A(L)$ increase with $L$, without saturating for the
values of $L$ available. This is consistent with the results
shown in Fig.~\ref{fig:bigfig} where two separate peaks in 
$g_s$ clearly are visible at large $L$ for these two values of
$t$. 

We have also tried to analyze the Thouless numbers by integrating
separately over the regions inside and outside the extended energies.
This is difficult to do for small $t$ since the extended energies 
cannot be located with a very high precision. Analyzing the integrated 
Thouless numbers separately for the two regions it is clear that
the number of extended states between the two extended energies
decreases significantly slower with $L$ than for the region
outside the extended energies.

\section{Network model}\label{sec:network} 

We now proceed to discuss our results for a network model of the
double-layer system. The network model was introduced by Chalker and
Coddington~\cite{ChalkCod} to take into account the corrections to
percolative behavior that occur when the correlation length
diverges in the vicinity of the extended state energies in
the middle of the Landau level. It is possible to map the network
model for a single layer on to various spin models~\cite{SUN,Ludwig}
and the calculated effective correlation lengths can be used
to estimate $\sigma_{xx}$ and $\sigma_{xy}$~\cite{LWK}.
The model that we use is essentially 
identical to one that has been studied in previous work 
by Lee et. al.~\cite{Derek}. Each individual layer is represented
by a separate network model in the manner described in
Ref.~\onlinecite{ChalkCod}.  A question now arises as to how to include
tunneling between the layers. We follow Ref.~\onlinecite{Derek} and
introduce a second saddle point along the straight paths in the
original network model, coupling the two layers. 
A priori there are several ways to represent such a saddle
point by a matrix. Since we want to model two
physical layers and not pseudospins we make a slightly different
choice than Ref.~\onlinecite{Derek}. We take the inter-layer
saddle point to be identical to the intra-layer saddle points,
i.e. represented by the following matrix:
\begin{equation}
{\bf T}_t=\left(\matrix{ {\bf M}_t^0 & 0 \cr
                        0 & {\bf M}_t^1  }\right),
\end{equation}
with ${\bf M}_t^n$ given by:
\begin{equation}
\left(\matrix{ e^{i\phi_{1}^n} & 0 \cr
	 0 & e^{i\phi_{2}^n} \cr  }\right)
\left(\matrix{ \cosh \theta_t & \sinh \theta_t \cr
	 \sinh \theta_t & \cosh \theta_t \cr  }\right)
\left(\matrix{ e^{i\phi_{3}^n} & 0 \cr
	 0 & e^{i\phi_{4}^n} \cr  }\right).
\label{eq:thetat}
\end{equation}
Here $\theta_t$ is the parameter that controls the tunneling between
the two layers and $n$ is the channel index counting the number of
channels in each layer. We shall always take $\theta_t$ to be constant. 
Random phases are included along the straight paths as in 
Ref.~\onlinecite{ChalkCod} described by the $\phi$'s in
Eq.~(\ref{eq:thetat}).
The
saddle points in the two layers are represented by identical
matrices but with a different parameter, $\theta$, which we again
take to be a constant and the {\it same} in the two layers.
\begin{equation}
{\bf T}_x=\left(\matrix{ \cosh \theta_a & 0 & \sinh \theta_a & 0\cr
                        0 & \cosh \theta_b & 0 & \sinh \theta_b\cr
			\sinh \theta_a & 0 & \cosh \theta_a & 0\cr
			0 & \sinh \theta_b & 0 & \cosh \theta_b\cr}\right).
\label{eq:thetax}
\end{equation}
Here the index $a,b$ refers to the two layers. We shall always take
$\theta_a=\theta_b$ since we are interested in modeling layers with
equal density.  

The choice of the matrix coupling the two layers, Eq.~(\ref{eq:thetat}),
is by no means obvious.
We could have used trigonometric functions instead of hyperbolics
as in Ref.~\onlinecite{Derek} thereby implying that in the picture
where an individual layer is represented by coupled
lines of opposite going currents (see for instance
Ref.~\onlinecite{ChalkDohm}) the two layers are {\it stacked in
register} with the currents going in the same direction in the two
layers. One can also stack the layers {\it out of register} and
the choice we have made corresponds in a certain sense
to a mixture of these two choices. We believe that none of these
microscopic details should matter for the universal properties of the
model and in particular for the divergence of the correlation length
at least for small $\theta_t$.
We have explicitly checked this for the results presented below in 
Fig.~\ref{fig:net0.05}, by repeating the calculation for several other
choices of stackings and coupling matrices. For the small coupling
of $\theta_t=0.05$, used in Fig.~\ref{fig:net0.05}, no dependence
on the microscopic results was observed.

We determine the correlation lengths associated with double-layer
systems, described by the transfer matrices outlined above, by
estimating the Lyaponov exponents.
The positive Lyaponov exponents, $\lambda^i_M(\theta)$, and their
uncertainties are calculated
following the method in Ref.~\onlinecite{KramerMac} for a range of
values of $\theta$ for fixed $\theta_t$. The correlation length
is determined as the inverse of the {\it smallest positive} Lyaponov
exponent, $\lambda^1_M(\theta)$. 
\begin{equation}
\xi_M(\theta)=1/\lambda^1_M(\theta).
\end{equation}
It is only necessary to calculate the positive Lyaponov exponents
which saves considerable computing time. An additional check on the
calculation can be done by calculating the first (and smallest in
absolute value) of the negative Lyaponov exponents which should be
the negative of $\lambda^1_M(\theta)$~\cite{D2}.

In general the correlation length, $\xi_M$, will be limited by the width
of the strip, $M$. The relevant quantity to study is therefore
the reduced correlation length, $\Lambda_M(\theta)=\xi_M/M$. An
insulating region will be characterized by $\Lambda_M(\theta)\to 0$,
a metallic one by $\Lambda_M(\theta)\to\infty$, or constant.
Since $\Lambda_M(\theta)$ is dimensionless, standard finite-size scaling
arguments predicts the scaling form:
\begin{equation}
\Lambda_M(\theta)=f(\xi/M),
\label{eq:fsz1}
\end{equation}
where $\xi$ is the correlation length in the infinite system.
As the critical energy, $E_c$, is approached this correlation length
diverges with the exponent $\nu$,
\begin{equation}
\xi \sim |E-E_c|^{-\nu}\equiv|\gamma|^{-\nu}.
\end{equation}
The relation between $\theta$ and the distance to the critical energy, 
$\gamma_c$, can be determined 
approximately~\cite{Fertig,Jaeger} for positive
$\theta$,
\begin{equation}
\gamma \approx \ln\sinh\theta, \ \theta > 0,
\label{eq:gamma}
\end{equation}
implying that the critical $\theta$ is given by $\sinh\theta_c=1$, or
$\theta_c=0.8814\ldots$. The relation Eq.~(\ref{eq:gamma}), allows
us to rewrite the finite-size scaling relation, Eq.~(\ref{eq:fsz1}),
in the following form,
\begin{equation}
\Lambda_M(\theta)=g(\gamma M^{1/\nu}),
\end{equation}
which is the form we shall use in the analysis of the numerical results.

\subsection{Computational Method}
We perform the calculations in a cylinder geometry imposing periodic
boundary conditions in the transverse direction. The system is made
invariant under a rotation by $90^o$ by alternating transfer
matrices with $\theta$ and
$\sinh\theta^\prime=1/\sinh\theta$ as described 
in Ref.~\onlinecite{ChalkCod}.
We denote the width of each of the two layers by $M$ and consider
systems with $M$ ranging from 4 to 128. For $M=4,8$ we generate
$2\times10^6$ transfer matrices and for the remaining widths
$2\times10^5$. We obtain sets of data by fixing $\theta_t$ and
approaching the critical point by varying $\theta$ and thereby $\gamma$.
As a check on our calculations we set $\theta_t$ to zero and
were able to reproduce the single-layer results from
Ref.~\onlinecite{ChalkCod} to within statistical errors. The integral
factor sometimes introduced in the definition of the number of layers
was chosen so that this would be the case.

\subsection{Results}
\begin{figure}
\centering
\epsfxsize=8.5 cm
\leavevmode
\epsffile{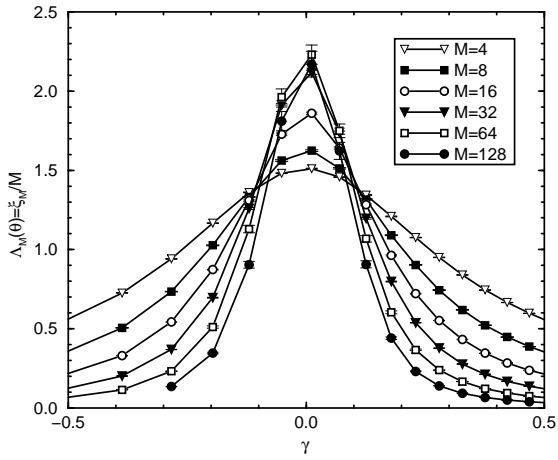}
\caption{The reduced correlation length as a function of the
energy $\gamma=\ln\sinh\theta$ for the double layer network
model with a tunneling parameter of $\theta_t=0.05$.}
\label{fig:net0.05}
\end{figure}

Our main results on the network model for the double-layer system 
is presented in Fig.~\ref{fig:net0.05}. For lattice sizes ranging
from $M=4$ to $M=128$ we have calculated the reduced correlation
length, $\Lambda_M(\theta)$. Since we want to view the tunneling
between the two layers, described by $\theta_t$, as a small perturbation
we take this parameter to be very small and constant, $\theta_t$=0.05.
Roughly we have the relation~\cite{ChalkDohm} $\tanh\theta_t= t$, where
$t$ is the parameter describing the tunneling in the exact
diagonalization studies in Section~\ref{sec:diags}. Note, that
we get $t$ and not $2t$ since we do not have periodic boundary
conditions between the two layers as we had in the
Section~\ref{sec:diags}. We then vary the intra-layer coupling,
$\theta$, and plot the results as a function of
$\gamma$.
As clearly seen in Fg.~\ref{fig:net0.05} we do not
reach a scaling regime until the width of the strips, $M$,
exceeds $1/t\sim 20$, as expected from the discussion in 
Section~\ref{sec:args}. We expect the extended state energi(es) to be
located at $\gamma\sim 0$ since we have taken the intra-layer
coupling, $\theta$, to be the same in the two layers. For 
widths larger than M=20 we observe that the reduced correlation length
becomes independent of the width, $M$, at $\gamma=0$ as expected.
For the sizes considered we do not see any signs of a splitting
of the two extended state energies which should have been of the
order of $\Delta\gamma\sim 2t\sim2\theta_t$,
based on the simple tight-binding picture, and
therefore clearly visible in Fig.~\ref{fig:net0.05}.

\begin{figure}
\centering
\epsfxsize=8.5 cm
\leavevmode
\epsffile{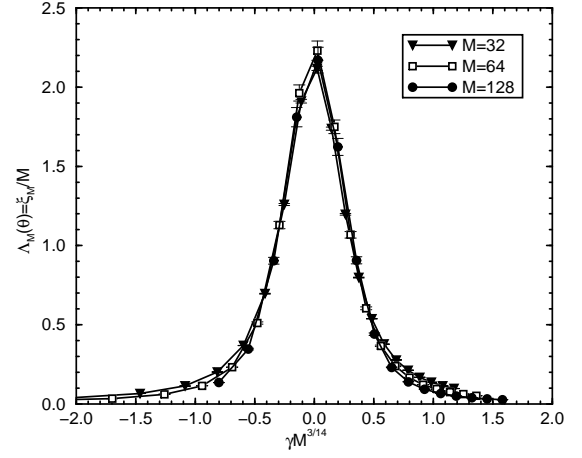}
\caption{Scaling collapse of the reduced correlation length 
for the double layer network model with a tunneling parameter
of $\theta_t=0.05$.}
\label{fig:netscale}
\end{figure}

Given these observations we therefore perform a scaling analysis
under the assumption that both of the two expected extended state
energies
are to be found at the same critical energy $\gamma_c=0$. Since we
do not expect scaling to be obeyed for $M\ll 20$ we only include
widths $M\gg 20$. Testing the scaling analysis is now a simple
matter of rescaling the $x$-axis in Fig.~\ref{fig:net0.05}
by an amount $M^{1/\nu}$ for the different widths. The result
is shown in Fig.~\ref{fig:netscale}.
Clearly very good scaling
is found for the chosen value of the correlation length exponent
$\nu=14/3$. Since we do not have a large number of numerical data
available to determine the exponent, $\nu$, we can only test
if a given value of $\nu$ gives good scaling. We tried $\nu=7/3$
and $\nu=11/3$ in both cases we found scaling that visibly was
much worse than what is shown in Fig.~\ref{fig:netscale} with
$\nu=14/3$. We therefore conclude that the apparent doubling of
the correlation length exponent is not in disagreement with
the numerical results obtained from the simplified network model
for the double-layer system.

\begin{figure}
\centering
\epsfxsize=8.5 cm
\leavevmode
\epsffile{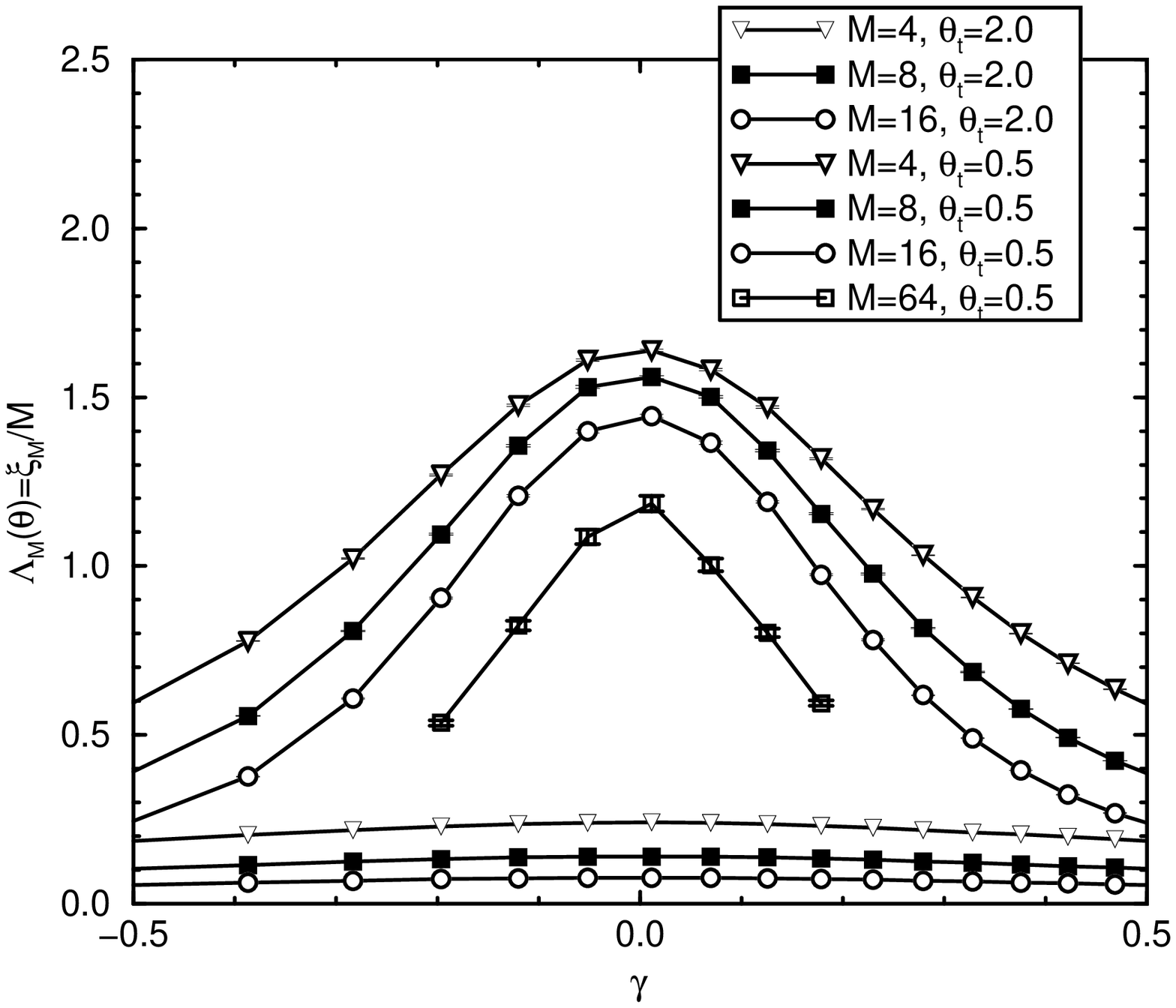}
\caption{The reduced correlation length for two different values
of $\theta_t=0.5$ and $2.0$.
the results are for for the double layer network model.}
\label{fig:netother}
\end{figure}

We now wish to make a few comments on the applicability of the network
model to the real physical system. We take the view that the starting
point for the network models is a percolation path close to a
critical energy. 
Quantum tunneling at the saddle points is then  
introduced as a {\it small} correction to the physics~\cite{Foot2}. 
We believe that the double-layer network
model only describes the physics of the coupled quantum wells
in detail,
in the limit where the inter-layer tunneling parameter tends to
zero, $\theta_t\to 0$. Our argument goes as follows: Imagine
we wanted to build a network model describing a {\it strongly
coupled} double-layer system. As we saw in Section~\ref{sec:diags}
the symmetric and antisymmetric states are then widely split, by
an amount $\Delta_{SAS}\sim 4t$. As $t$ becomes very large we
should therefore model the system as two {\it weakly coupled}
single layer networks, describing the symmetric and antisymmetric
states. These two networks should then be modeled using different
interlayer couplings $\theta_1$ and $\theta_2$ corresponding to
the symmetric and antisymmetric states.
In the limit $t\to\infty$ the two networks become completely
decoupled (and $|\theta_1-\theta_2|\to\infty$)
and we should thus set $\theta_t=0$. Related ideas were
proposed for a network model in Ref.~\onlinecite{Shahbazyan} 
in order to describe Landau level mixing 
and related ideas have also previously been
discussed in Ref.~\onlinecite{WLW}.
In our model, as we have
described it above, the two networks correspond to the two layers
in physical space. As the intra-layer coupling is increased we
{\it do not form symmetric and antisymmetric states}, but rather
the paths become localized in orbits between the two layers.
This is clearly seen in Fig.~\ref{fig:netother} where we plot results
for $\theta_t=0.5$ and $\theta_t=2.0$. 
Clearly, the reduced correlation
length is growing slower than $M$, and for $\theta_t=2.0$ it
appears that it is independent of $M$ and also of $\gamma$, consistent
with our expectation that the orbits all are localized between
the two layers on a length scale of order 1. This is supported
by the observation that for $\theta_t=2.0$,
$\Lambda_M(\theta)$
{\it decreases} with $M$ roughly 
as $M^{-1}$ indicating a constant correlation
length. For intermediate couplings it is possible that one could
see structure resembling the symmetric and antisymmetric state
but we believe that the double-layer network models only describe
the correct physics of the real double-layer systems in the
limit where the inter-layer tunneling parameter, $\theta_t$, is
but a small perturbation. Hence, we believe that the double-layer
network model do not describe the microscopic physics in detail
for intermediate tunneling parameters.

\section{Summary and Discussion}\label{sec:conclusion}

We have shown that in the strong magnetic field limit non-interacting
double-layer electron systems in which interlayer tunneling occurs have
two extended state energies for each orbital Landau level.  Numerical
results show that, for large tunneling amplitude, Landau levels
associated with subbands which are symmetric and antisymmetric
combinations of isolated layer states are weakly mixed by disorder.
Localization properties within the Landau levels of the symmetric and
antisymmetric subbands are similar to those for a single 2D electron
layer and in particular appear to have a correlation length exponent,
$\nu$, identical to what was found for an isolated layer to within
numerical precision.   For smaller values of the tunneling amplitude
where symmetric and antisymmetric subbands are not well developed in the
density of states, numerical results still appear to show that extended
states occur at only two energies which are split by an amount somewhat
larger than the splitting of symmetric and antisymmetric Landau levels
in the absence of disorder.  We cannot exclude the possibility that a
finite amount of tunneling is necessary to split the two extended state
energies, although the weight of available evidence appears to suggest
the contrary.  However, numerical values of the localization length are
much larger than in the limits of either strictly zero tunneling or
large tunneling.  This is especially true in the energy interval between
the two extended state energies.  Over the range of system sizes
accessible for numerical studies the exponent for the diverging
localization length appears to be approximately twice as large as for
the case of isolated layers.  We have compared these microscopic
calculations with network models of double-layer systems.  The network
model and microscopic numerical results differ qualitatively.  In the
network model case there is no evidence for two discrete critical
energies at which extended states occur.  We conclude from our study
that, in contrast with the single layer case, network models do not
generically give reliable results for the strong magnetic field
localization properties of double-layer systems.

We believe that our results have important implications for the integer
quantum Hall effect in three-dimensional electron systems.   We comment
here only on the case where the band width along the field direction is
smaller than the Landau level separation and the quantum Hall effect has
the best chance of occurring.  It is important to realize that the
physics of this extreme strong field regime is qualitatively different
from the more usual three-dimensional case where many different Landau
`tubes' cross~\cite{schoenberg} the Fermi energy.   (The physical
systems we have in mind are multiple quantum well (MQW) systems, like
those studied experimentally by St\" ormer {\it et al.}, with weak
barriers between the wells.)   In particular, just as in the single
layer integer quantum Hall case, we can argue that disorder can never
result in the localization of all states.  This point is perhaps made
most elegantly using the topological picture of the integer quantum Hall
effect\cite{stonesbook}.  In the absence of disorder an elementary
calculation shows that the Hall conductance in $e^2/h$ units, and hence
the sum of the Chern numbers of all states, is equal to the number of
layers in the MQW.  As the states evolve adiabatically with disorder,
the sum of all the Chern numbers of all states in the (energy range of
interest) cannot change\cite{stonesbook}.  Since only extended states
can have non-zero Chern numbers, it is impossible to localize all
states.  The situation is closely analogous to the quantum Hall effect
in a single two-dimensional electron gas in a magnetic field when Landau
level separations become small, since each Landau level contributes $1$
to the Chern number sum and localization is possible only by mixing
states with different Chern number which are at energies well away from
the bottom of the band of the host semiconductor.

In the absence of disorder the states in the energy range of interest in
the MQW consist of a set of macroscopically degenerate Landau levels,
labeled by wave-vectors and split by an amount proportional to the
interlayer hopping amplitude.  For small $\Gamma/t$, states with a given
wave-vector will be weakly coupled and a single extended state energy
will exist for each Landau level.  As the disorder strength increases,
our numerical results for the two layer case suggest that the extended
state energies will remain separate but localization lengths will
increase substantially, except at energies above the highest extended
state energy and below the lowest extended state energy.  (However, we
know of no general argument which forbids either the collapse of the
extended states toward a single energy or, in the other extreme, the
development of a band of energies over which states are extended.)  In
the limit of an infinite number of layers, the energy separation between
extended states will approach zero but the system will still, strictly
speaking, not be metallic since almost all states in any range will
still be localized~\cite{Foot3}.  Our numerical results suggest the
possibility that critical exponents for localization lengths diverging
between intermediate energy extended state could be different from the
critical exponents for single-layer systems.  

It is interesting to consider whether or not the metallic phase of
three-dimensional systems in a strong magnetic field suggested by the
work in Refs.~\onlinecite{Ohtsuki1,Henneke,ChalkDohm}, can be reconciled
with the expectation from integer quantum Hall theory and from the
present calculations of a discrete set of extended state energies for
any finite number of layers.  In order for these two pictures to be
compatible the localization lengths at energies between the lowest and
highest extended state energies would have to increase with the number
of layers and diverge in the limit of infinite layer numbers or
equivalently the limit of small separations between extended state
energies.  Incidentally, if we assume that each additional quantum well
leads to another extended state energy {\it without} changing the
associated correlation length exponents as this energy is approached, it
is not clear how to explain the large difference between the exponent
$\nu$ found at the mobility edge in
Refs.~\onlinecite{Ohtsuki1,Henneke,ChalkDohm} and the two-dimensional
exponent.  We believe that the tendency toward an apparent metallic
phase in multiple quantum well systems in the limit of large layer
numbers or small interlayer hopping is extremely closely connected with
the disappearance of the quantum Hall effect in a high-mobility
two-dimensional electron system in the limit of weak magnetic fields
since disorder in both cases permits only mixing of Landau bands
carrying the same unit total Chern number.  Existing experiments on the
integer quantum Hall effect in MQW systems have observed a quantized
Hall effect only at Fermi energies above the highest energy extended
state where electrons are well localized.  We hope that the present
paper will motivate new attempts to study the physics of the quantum
Hall effect in double quantum well and MQW systems at Fermi energies
between extended state energies.  

\acknowledgments

We gratefully acknowledge discussions with L.~Balents, S.~Cho, 
M.~P.~A.~Fisher, S.~M.~Girvin, C.~B.~Hanna, D.~K.~K.~Lee, and J.~J.~Palacios. 
This research is supported by NSF grant number NSF DMR-9416906.

\begin{references}
\bibitem{Foot1}
This finding suggests that interaction effects must be important
at some level since one would expect $z=2$ for truly non-interacting
electrons.
\end{references}
\end{document}